\documentclass{PoS}
\usepackage{mathtools}
\title{Recent advances on chiral partners and patterns and the thermal $f_0(500)$ in chiral restoration}

\ShortTitle{Recent advances on chiral partners and patterns}

\author{\speaker{Angel G\'omez Nicola}
\\
        Departamento de F\'{\i}sica
Te\'orica and UPARCOS. Univ. Complutense. 28040 Madrid. Spain\\
        E-mail: \email{gomez@ucm.es}}

\author{Jacobo Ruiz de Elvira\\
        Albert Einstein Center for Fundamental Physics, Institute for Theoretical Physics,
University of Bern, Sidlerstrasse 5, CH--3012 Bern, Switzerland\\
        E-mail: \email{elvira@itp.unibe.ch}}

\author{Silvia Ferreres-Sol\'e\\
     Department of Astronomy and Theoretical Physics,\\
Lund University, SE-223 62 Lund, Sweden
\\
        E-mail: \email{ferreres.sole@gmail.com}}
\author{Andrea Vioque-Rodr\'iguez\\
         Departamento de F\'{\i}sica
Te\'orica. Univ. Complutense. 28040 Madrid. Spain\\
        E-mail: \email{avioque@ucm.es}}
        
\abstract{We review recent results on chiral $SU(2)_L\times SU(2)_R\approx O(4)$ and $U(1)_A$ symmetry restoration in QCD. 
In particular, we discuss how  Ward Identities allow one to derive general results on partner degeneration, which shed light on the distinction between the $O(4)$ and $O(4)\times U(1)_A$ patterns of the chiral transition.
For that purpose,  susceptibilities associated with the $O(4)$ and $U(1)_A$ symmetries are studied. 
From this analysis we conclude that in the ideal regime of exact $O(4)$ restoration (formally achieved in the limit of two massless flavours), $U(1)_A$ partners degenerate as well. 
We also discuss the role of the thermal $f_0(500)$ state to describe thermodynamic observables sensitive to chiral restoration, such as the scalar susceptibility. 
We pay special attention to the consistency of our results with recent lattice analysis.}

\FullConference{XVII International Conference on Hadron Spectroscopy and Structure - Hadron2017\\
		25-29 September, 2017\\
		University of Salamanca, Salamanca, Spain}

\newcommand{\intT}{\int_0^\beta d\tau \int d^3 \vec{x}}
\newcommand{\eqchiral}{\!\stackrel{O(4)}{\sim}\!}

\newcommand{\condl}{\mean{\bar q q}_l}
\newcommand{\conds}{\langle \bar s s \rangle}
\newcommand{\mean}[1]{\left\langle{#1}\right\rangle}
\newcommand{\quarkcorT}{\langle {\cal T} (\bar \psi_l\psi_l)(x) \,(\bar \psi_l
 \psi_l)(0)\rangle_{T}}
\newcommand{\im}{\mbox{Im}\,}

\begin{document}
\section{Introduction}
Chiral symmetry restoration is one of the main topics regarding the understanding of the QCD phase diagram. 
Being linked to deconfinement, at least from the viewpoint of the coincidence of the transition region, its properties and nature constitute a very active topic of study for current theoretical, phenomenological and lattice analysis. 
Its experimental implications are also crucial for the Physics of Ultrarelativistic Heavy Ion Collisions. It has been already established that in the physical case of $N_f=2+1$ flavours with $\hat m\ll m_s$  masses, the chiral transition is a  crossover at a transition temperature of about $T_c\sim 155-160$ MeV for vanishing baryon density~\cite{Aoki:2009sc,Bazavov:2011nk,Buchoff:2013nra}. The ideal chiral restoration phase transition is reached only for $N_f=2$ and $\hat m =0$, while in the physical case it is approached in the light chiral limit $\hat m\rightarrow 0^+$. A particularly relevant question in this context is whether the asymptotic restoration of the $U(1)_A$ symmetry~\cite{Gross:1980br} can take place effectively already at the $O(4)$ transition. If that was the case, the chiral pattern would become $O(4)\times U(1)_A$, which would have different theoretical and phenomenological implications including the transition order~\cite{Pisarski:1983ms} or the dilepton and photon spectrum modification from the $\eta'$ mass reduction~\cite{Kapusta:1995ww,Csorgo:2009pa}. 

This problem has been recently investigated by different lattice collaborations, mostly through the study of chiral and $U(1)_A$ partners, i.e, states becoming degenerate under those symmetry groups. 
Here, we consider the members of the pseudoscalar and scalar nonets,  $\pi^a=i\bar\psi_l\gamma_5\tau^a\psi_l$, $\delta^a=\bar\psi_l \tau^a \psi_l$ , $\eta_l=i\bar\psi_l\gamma_5 \psi_l$, $\eta_s=i\bar s \gamma_5 s$, $\sigma_l=\bar\psi_l \psi_l$, $\sigma_s=\bar s  s$, $K^a=i\bar\psi  \gamma_5 \lambda^a \psi$, $\kappa^a=i\bar\psi  \lambda^a \psi$, with $\psi_l$  the light quark doublet, which correspond to the quantum numbers of the pion, $a_0(980)$, light and strange component of the $\eta/\eta'$, light and strange components of the $f_0(500)/f_0(980)$, kaon and $K(800)$ (or $\kappa$), respectively. 
For the isospin $I=0,1$ sector, chiral and $U(1)_A$ transformations connect the bilinears $\pi^a\,\xleftrightarrow{SU_A(2)}\,\sigma$, $\delta^a\xleftrightarrow{SU_A(2)}\eta_l$, $\pi^a\xleftrightarrow{U(1)_A}\delta^a$ and $\sigma\xleftrightarrow{U(1)_A}\eta_l$.

\begin{figure}
\centerline{\includegraphics[width=9cm]{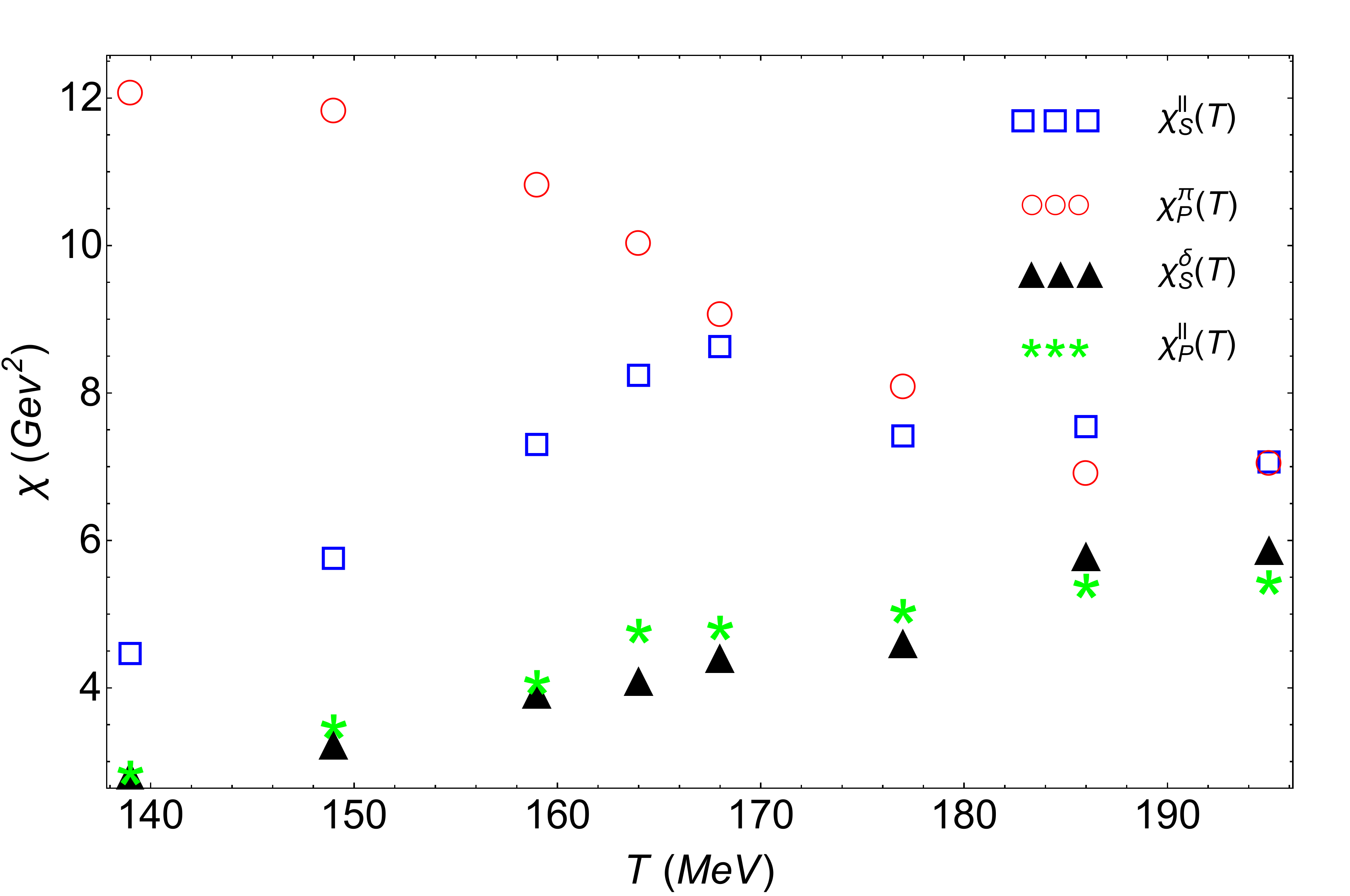}}
\caption{Lattice data ~\cite{Buchoff:2013nra} ($32^3\times 8$ lattice size) for  susceptibilities  in  $O(4)$ and $U(1)_A$ restoration.}
\label{fig:latticenonet}
\end{figure}

The study of the chiral symmetry pattern through the degeneration of the above partners is customarily done through the susceptibilities of the corresponding correlators, defined as:
\begin{equation}
 \chi_Y(T)=\int_T dx  \langle \mathcal{T} Y(x) Y (0) \rangle,
\end{equation}
with $\int_T dx\equiv\intT$ and $Y=P,S$ the pseudoscalar and scalar quark bilinears defined above. 
Susceptibilities correspond to the $p=0$ correlators in momentum space and for the particular cases $Y=\bar \psi_l\psi_l$ and $Y=\bar s s$,  subtracting  $\int_T dx \langle Y\rangle^2$, 
one gets the mass derivative of the corresponding quark condensate~\cite{GomezNicola:2012uc}. 
Note in particular that the susceptibility difference $\chi_{5,disc}=\frac{1}{4}\left(\chi_P^\pi-\chi_P^{ll}\right)$ is a measure of the $O(4)\times U(1)_A$ breaking. That particular combination is also interesting because it is proportional to the topological susceptibility measuring genuine anomalous effects~\cite{Buchoff:2013nra}.  Actually, 
in~\cite{Buchoff:2013nra}, for $N_f=2+1$ and physical quark masses, a sizable separation between the susceptibilities for the $O(4)$ and $U(1)_A$ partners was obtained. 
We plot their results in Fig.~\ref{fig:latticenonet}. However, another lattice group~\cite{Aoki:2012yj,Cossu:2013uua,Tomiya:2016jwr} has obtained results compatible with the $O(4)\times U(1)_A$ pattern at chiral restoration, for $N_f=2$ in the chiral limit. Furthermore, in~\cite{Brandt:2016daq}, the  same $O(4)\times U(1)_A$ pattern is also obtained for $N_f=2$ and nonzero quark masses. 

Here, we will review our recent analysis ~\cite{Nicola:2016jlj,GomezNicola:2017bhm} of Ward Identities (WI) in connection with the above pattern and partner problem. In addition, we will discuss how the thermal $f_0(500)$ pole plays a crucial role to describe correctly the  chiral susceptibility in the scalar channel. 
\section{Ward Identities: chiral partners and patterns}
In order to shed light on the above chiral pattern problem, we have studied~\cite{GomezNicola:2017bhm} a set of WI derived formally from QCD, 
which allows one to connect susceptibilities with quark condensates and differences of susceptibilities with three-point meson vertices.  
A particular relevant identity can be obtained by combining the pseudoscalar WIs  derived in~\cite{Nicola:2016jlj}:
\begin{equation}
\chi_P^{ls}(T)=-2\frac{\hat m}{m_s} \chi_{5,disc}(T)
\label{wils5}
\end{equation}
where $\chi_P^{ls}$ is in general nonzero due to $\eta/\eta'$ mixing and $\hat m \chi_{5,disc}$ remains nonzero in the light chiral limit at $T=0$. 
The importance of~\eqref{wils5} arises from the fact that if $O(4)$ partners are degenerated $\chi_P^{ls}\eqchiral 0$ since a $O(4)$ transformation rotates $\eta_l\rightarrow \delta^a$ but leaves $\eta_s$ invariant, 
so that $\chi_P^{ls}$ rotates into a correlator which vanishes by parity. Thus,~\eqref{wils5}) implies that in such $O(4)$ degenerating scenario, $\chi_{5.disc}$ (and hence the topological susceptibility) should vanish as well, 
which supports the $O(4)\times U(1)_A$ pattern from the point of view of partner degeneration. This conclusion is consistent with the lattice works~\cite{Aoki:2012yj,Cossu:2013uua,Tomiya:2016jwr,Brandt:2016daq}, 
while the discrepancy with~\cite{Buchoff:2013nra} can be explained~\cite{GomezNicola:2017bhm} by the large uncertainties in the degeneration of $\chi_P^{ll}-\chi_S^\delta$ compared to those in $\chi_P^{\pi}-\chi_S^{ll}$,  
visible in Fig.~\ref{fig:latticenonet}.

Another consequence of the WI studied in~\cite{GomezNicola:2017bhm} is that the susceptibilities associated to the $K$ and $\kappa$ can be related in the following way:
\begin{equation}
  \chi_S^\kappa (T)-\chi_P^K (T)=\frac{2}{m_s^2-\hat m^2}\left[m_s\condl (T)-2\hat m \conds (T)\right],
  \label{wichikappakaondif}
\end{equation}
which implies that $K-\kappa$ also become degenerate in the ideal $O(4)$ restoration regime, i.e for $\hat m\rightarrow 0^+$ and $\condl\rightarrow 0^+$. 
Furthermore,~\eqref{wichikappakaondif} allows one to parametrize the breaking of that degeneracy in the lattice, since the r.h.s is proportional to the subtracted quark condensate $\Delta_{l,s}=\condl-2(\hat m/m_s)\conds$, 
customarily used in lattice as a chiral order parameter in order to avoid finite size divergences in individual condensates~\cite{Aoki:2009sc,Bazavov:2011nk,Buchoff:2013nra}.

\begin{figure}
\centerline{\includegraphics[width=9cm]{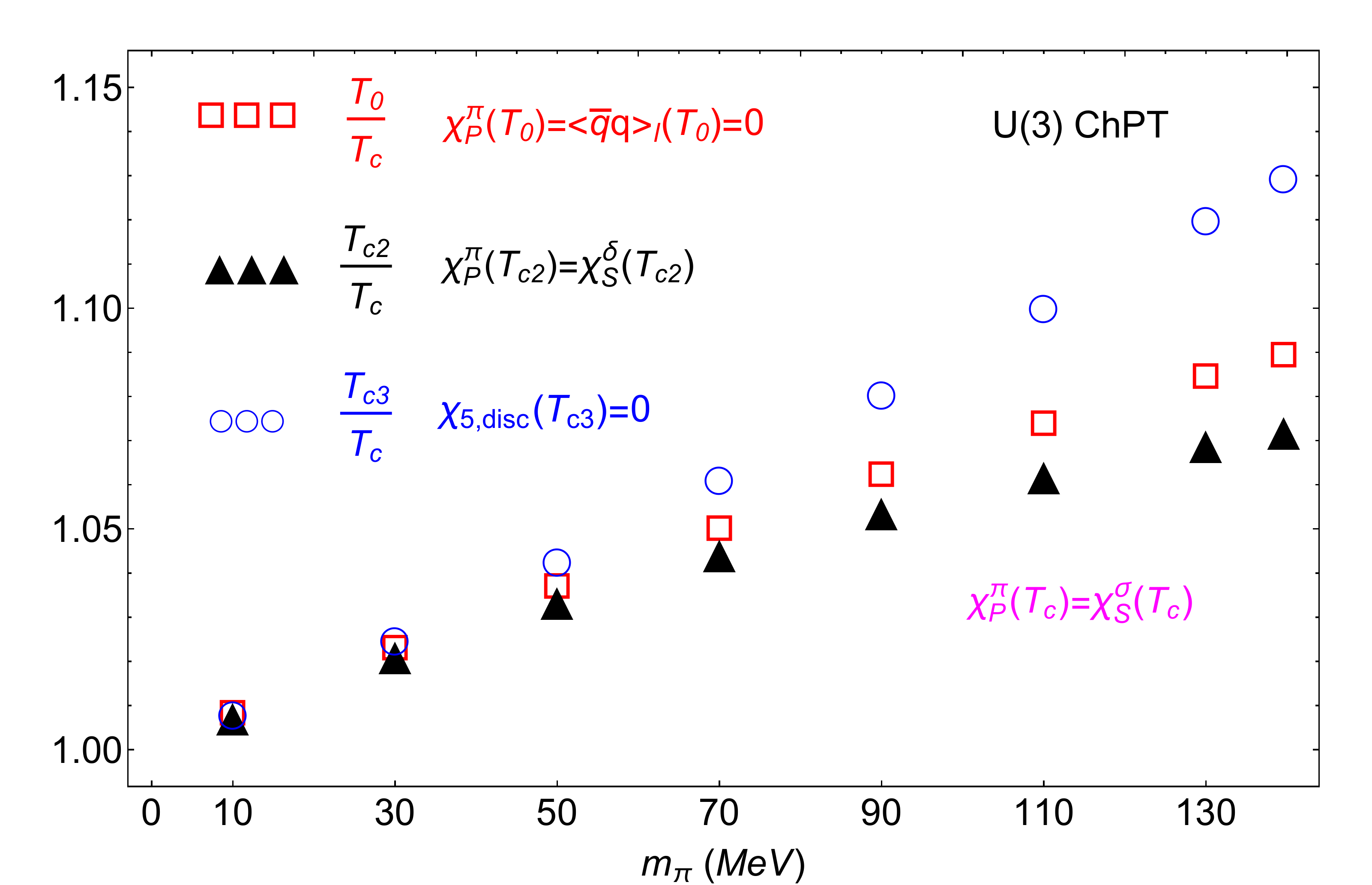}}
\caption{$U(3)$ ChPT results for the different partner degeneration temperatures}
\label{fig:chpt}
\end{figure}

We remark that the previous WI have been derived formally from QCD and hence they are model independent. 
Although the arguments used rely on symmetry properties of the operators involved, in general one must be careful about possible renormalization issues of those operators in QCD. 
For that purpose, we have checked their validity within  $U(3)$ Chiral Perturbation Theory (ChPT) \cite{Nicola:2016jlj,AGNJREnext}, which provides a model-independent and renormalizable set-up. The ChPT analysis also confirms our previous findings, in particular the coincidence of $O(4)$ and $U(1)_A$ partner degeneration in the chiral limit, for which we show preliminary results in Fig.~\ref{fig:chpt}. 

\section{The thermal $f_0(500)$}
As shown in the previous sections, the scalar susceptibility
\begin{equation}
\chi_S (T)=-\frac{\partial}{\partial m_l} \condl(T)=\int_E{d^4x \left[\quarkcorT-\condl^2\right]},
\end{equation}
plays a crucial role to describe chiral symmetry restoration. 
Not only it provides information about partner degeneration, but it is also a quantity directly sensitive to the transition, since it is expected to diverge for a second order phase transition~\cite{Smilga:1995qf}. 
In the physical case with $N_f=2+1$ flavors and massive quarks, the peak observed for $\chi_S$ in the lattice signals the transition temperature~\cite{Aoki:2009sc,Bazavov:2011nk,Buchoff:2013nra} as it is seen for $\chi_S^{ll}$ in Fig.~\ref{fig:latticenonet}.  

\begin{figure}
\centerline{\includegraphics[width=7.5cm]{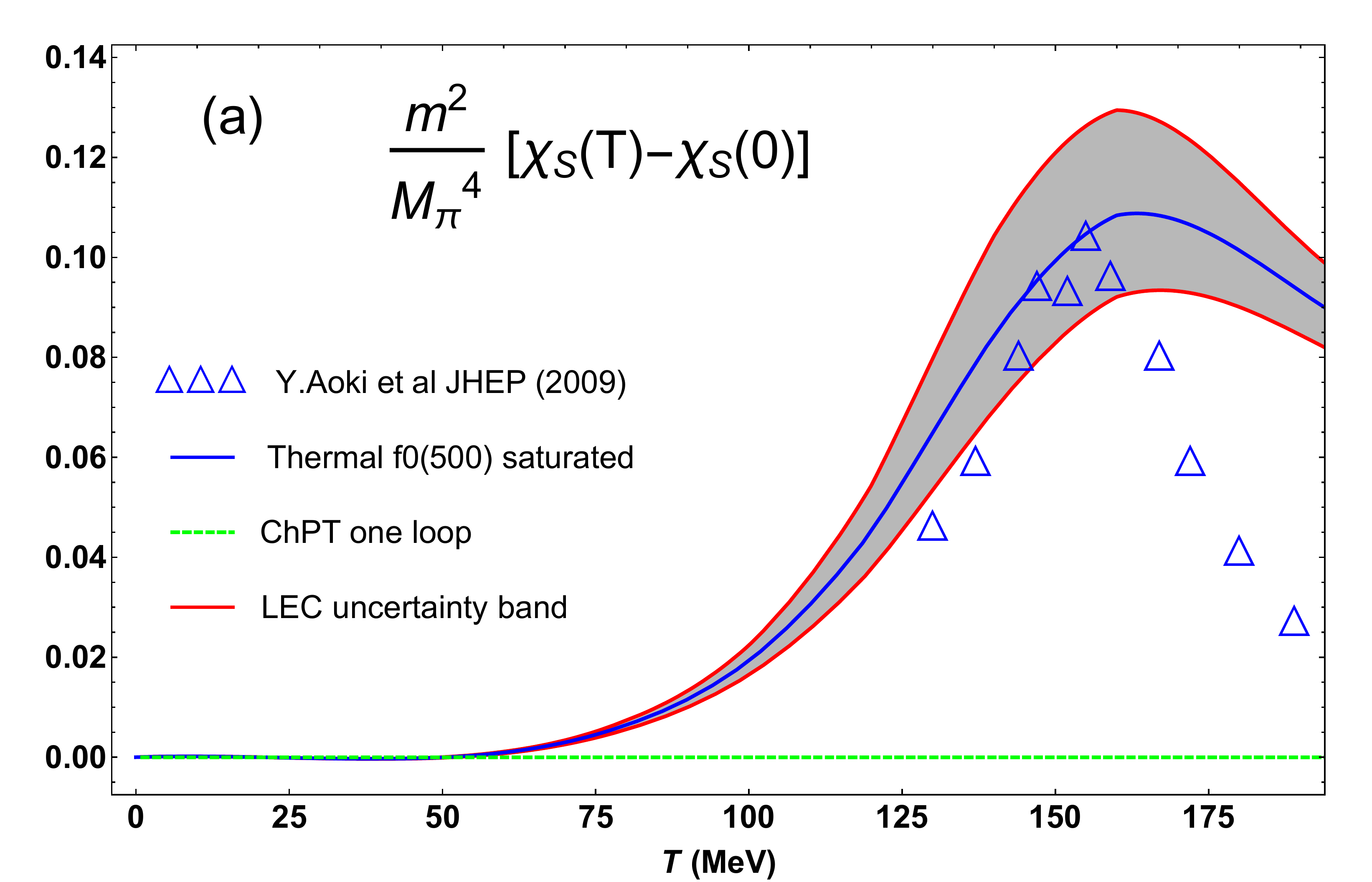}\includegraphics[width=7.5cm]{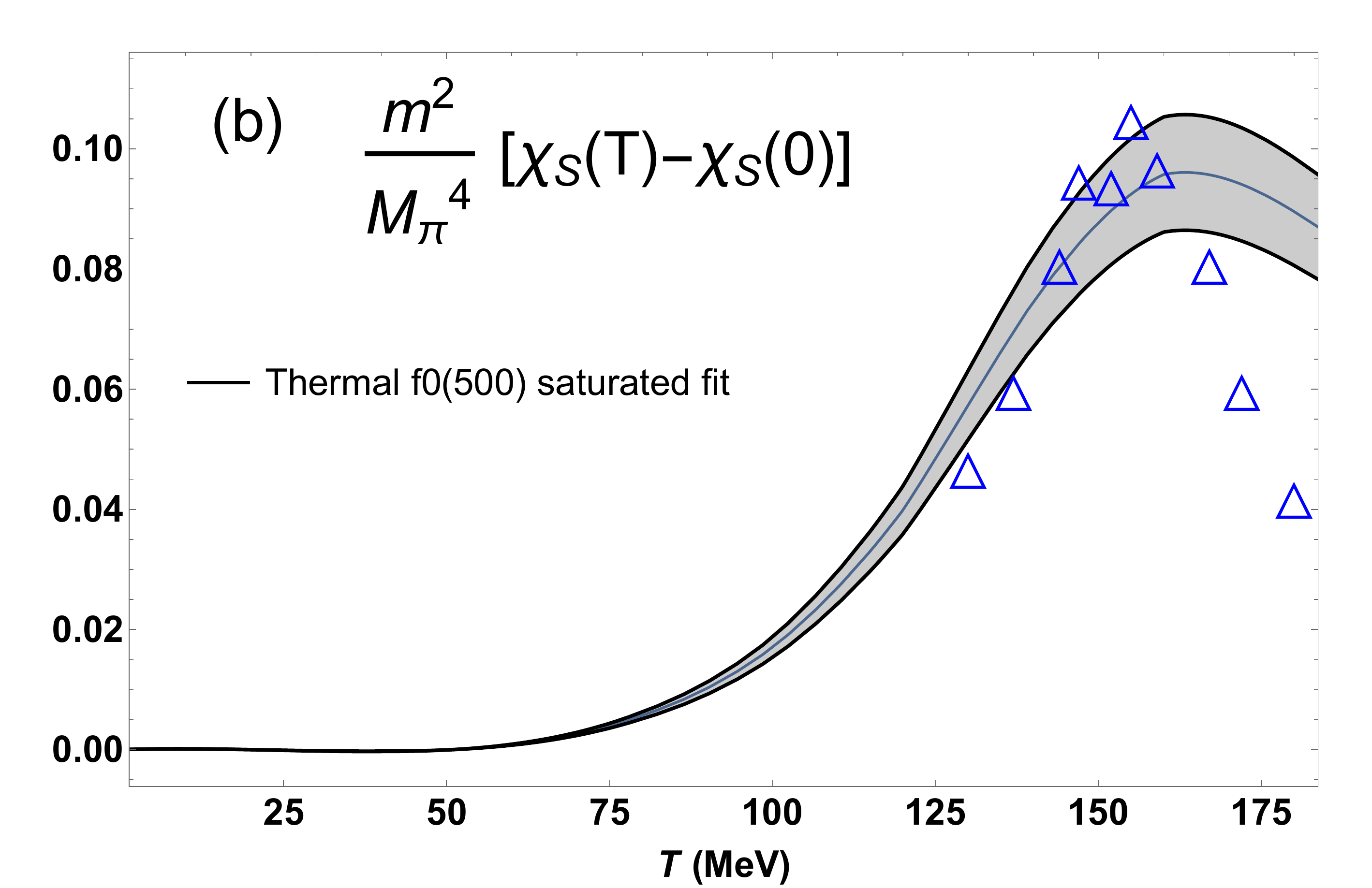}}
\caption{(a) Scalar susceptibility saturated by the  thermal $f_0(500)$ result as given in (3.3), compared with the ChPT result and with the lattice data in~\cite{Aoki:2009sc}. We include the uncertainty band coming from the LEC in ~\cite{Hanhart:2008mx}. (b) Fit of the unitarized $\chi_S$ to the latice data.}
\label{fig:suscep}
\end{figure}

On the one hand, the behaviour of $\chi_S(T)$ in hadron models such as ChPT shows a continuously increasing function of temperature~\cite{GomezNicola:2012uc}. However, the expected properties of the susceptibility can be recovered by recalling that it corresponds to the $p=0$ $\sigma$-channel propagator and then it is expected to behave, up to coupling normalizations, as $\chi_S\sim 1/M_S^2$ with $M_S$ the scalar mass of that channel, whose lightest hadron state is the $f_0(500)$.  On the other hand, it is already well established that the $f_0(500)$ pole can be generated, among  other techniques available in the literature~\cite{Pelaez:2015qba}, by constructing  a unitarized $\pi\pi$ scattering amplitude via the so called Inverse Amplitude Method 
\begin{equation}
t^{IAM}(s;T)=\frac{t_2(s)^2}{t_2(s)-t_4(s,T)}.
\label{iam}
\end{equation}
where $t_2+t_4+\dots$ is the ChPT series for the $I=J=0$ partial wave and the temperature dependence enters in $t_4$ through loop corrections as calculated in~\cite{GomezNicola:2002tn}. The amplitude~\eqref{iam} satisfies exactly the thermal unitarity relation $\im t^{IAM}(s+i\epsilon;T)=\sigma_T(s) \vert t^{IAM}(s)\vert^2$ ($s\geq 4M_\pi^2$) with 
$$\sigma_T (s)=\sqrt{1-4M_\pi^2/s}\left[1+2n_B(\sqrt{s}/2;T)\right],$$
the thermal two-particle phase space and $n_B(x;T)=\left[\exp(x/T)-1\right]^{-1}$ the Bose-Einstein distribution function.  In addition, the unitarized amplitude is analytic and hence the thermal $f_0(500)$ shows up as a pole in the second Riemann sheet at $s=s_p (T)=\left[M_p(T)-i\Gamma_p(T)/2\right]^2$~\cite{Dobado:2002xf}. The real part of the pole, which would correspond to the  self-energy real part of a scalar particle exchanged between the incoming and outgoing pions, defines a thermal mass for this state, namely $M_S^2(T)=M_p^2 (T)-\Gamma_p^2(T)/4$, which shows a clear dropping behaviour with $T$~\cite{Nicola:2013vma},  corresponding to the expected chiral restoration features for that state. From the previous arguments, it has been showed~\cite{Nicola:2013vma}  that defining the unitarized susceptibility as
\begin{equation}
\chi_S^U(T)=A\frac{M_\pi^4}{4\hat m^2}\frac{M_S^2(0)}{M_S^2(T)},
\label{susunit}
\end{equation}
and choosing $A=\frac{4\hat m^2}{M_\pi^4}\chi^{ChPT}_S (0)\simeq 0.14$ to match the ChPT result, one gets a peak very close to the expected $T_c$ lattice value. Roughly speaking, using (\ref{susunit}) we are assuming that the thermal $f_0(500)$ saturates the scalar susceptibility and that the possible $T$ dependence of the  $A$ factor (in particular from the effective $\sigma\pi\pi$ coupling residue) is smooth enough to be included in the uncertainties of the method.  Actually, the result is  very robust under variations in the unitarization method and under the uncertainty of the Low Energy Constants (LEC) involved~\cite{AGNFSVRnext}. We show in Fig.~\ref{fig:suscep}(a) preliminary results for this $f_0(500)$  saturated susceptibility, where the thermal pole has been calculated with the LEC given in~\cite{Hanhart:2008mx} including the LEC uncertainties quoted in that work. Just within the LEC uncertainty band, i.e without performing any fit to lattice $\chi_S$ data, this approach describes remarkably well the lattice results. 
A thermal model fit for the scalar susceptibility only in terms of such saturated thermal $f_0(500)$ gives also a very good description~\cite{AGNFSVRnext}. In Fig.~\ref{fig:suscep}(b) we show preliminary results of a fit using $A$ in~(\ref{susunit}) as fit parameter, also with the LEC in~\cite{Hanhart:2008mx}. The results in that figure correspond to fitting the lattice points up to $T=167$ MeV. The bands correspond to the 95\% confidence level interval of the fit, giving for $A=0.13\pm 0.01$ and hence compatible with the ChPT central value used in Fig.~\ref{fig:suscep}(a). For that fit we get  a coefficient of determination $R^2=0.988$. 
\section{Conclusions} 
Ward Identities are a powerful theoretical method, which allow one to reach relevant conclusions about chiral partners and the corresponding pattern. 
In particular, in the limit of ideal $O(4)$ partner degeneration, they also predict the degeneration of $U(1)_A$ partners, while $K-\kappa$ become ideal chiral partners in the $I=1/2$ sector. These results are confirmed by a $U(3)$ ChPT analysis. We have also showed recent results on the role of the thermal $f_0(500)$ pole for chiral restoration. That state generated within unitarized ChPT allows to describe the scalar susceptibility lattice data by assuming saturation with the associated thermal mass. 
\section*{Acknowledgments}
Work partially supported by  research contracts FPA2014-53375-C2-2-P, FPA2016-75654-C2-2-P and the Swiss National Science Foundation.

\end{document}